\documentclass[aps,prc,showpacs,superscriptaddress,byrevtex]{revtex4-1}  
\usepackage{amsmath}
\usepackage{color}
\usepackage{graphicx}
\usepackage{float}
\usepackage{ulem}
\begin{document}
\title{Investigation of the pairing effect in $^{10}$B nucleus compared with
$^{10}$Be and $^{10}$C nuclei by using the extended THSR wave function}

\author{Qing Zhao} \email{zhaoqing91@outlook.com.}
\affiliation{School of Physics and Key Laboratory of Modern Acoustics,
Institute of Acoustics, Nanjing University, Nanjing 210093, China}

\author{Zhongzhou Ren} \email{corresponding author: zren@tongji.edu.cn.}
\affiliation{School of Physics Science and Engineering,
Tongji University, Shanghai 200092, China}

\author{Mengjiao Lyu} \email{mengjiao@rcnp.osaka-u.ac.jp}
\affiliation{Research Center for Nuclear Physics (RCNP), Osaka
  University, Osaka 567-0047, Japan}

\author{Hisashi Horiuchi} \affiliation {Research Center for Nuclear
  Physics (RCNP), Osaka University, Osaka 567-0047, Japan}
\affiliation {International Institute for Advanced Studies, Kizugawa
  619-0225, Japan}

\author{Yoshiko Kanada-En'yo} \email[]{yenyo@ruby.scphys.kyoto-u.ac.jp}
\affiliation{Department of Physics, Kyoto University, Kyoto 606-8502, Japan}
  
\author{Yasuro Funaki} \affiliation{Laboratory of Physics, Kanto
  Gakuin University, Yokohama 236-8501, Japan}

\author{\mbox{Gerd R\"{o}pke}} \affiliation{Institut f\"{u}r Physik,
  Universit\"{a}t Rostock, D-18051 Rostock, Germany}

\author{Peter Schuck} \affiliation{Institut de Physique Nucl\'{e}aire,
  Universit\'e Paris-Sud, IN2P3-CNRS, UMR 8608, F-91406, Orsay,
  France} \affiliation{Laboratoire de Physique et Mod\'elisation des
  Milieux Condens\'es, CNRS-UMR 5493, F-38042 Grenoble Cedex 9,
  France}

\author{Akihiro Tohsaki} \affiliation{Research Center for Nuclear
  Physics (RCNP), Osaka University, Osaka 567-0047, Japan}

\author{Chang Xu} \affiliation{School of Physics and Key Laboratory of
  Modern Acoustics, Institute of Acoustics, Nanjing University,
  Nanjing 210093, China}

\author{Taiichi Yamada} \affiliation{Laboratory of Physics, Kanto
  Gakuin University, Yokohama 236-8501, Japan}

\author{Bo Zhou} \affiliation{Institute for International Collaboration, Hokkaido University, Sapporo 060-0815, Japan}
\affiliation{Department of Physics, Hokkaido University, 060-0810 Sapporo, Japan}


\begin{abstract}
In order to study the nucleon-nucleon pairing effects in clustering nuclei, we
formulate a superposed Tohsaki-Horiuchi-Schuck-R\"{o}pke (THSR) wave function,
which includes both molecular-orbit and pairing configurations
explicitly. With this new wave function, we investigate the abnormal
deuteron-like $pn$-pairing effect in $^{10}$B with T=0 and S=1 (isoscalar) by
comparing with isovector $NN$ pairs ($T=1, S=1$) in $^{10}$Be and $^{10}$C.
Energies are calculated for the ground states of $^{10}$Be, $^{10}$B and
$^{10}$C nuclei, and the $1^+_10$ excited state of $^{10}$B. These energies are
essentially improved comparing with studies using previous version of THSR wave
function. Further more, overlaps between the total wave function and the pairing
component indicate that the $NN$ pairing effect is more visible in $^{10}$B than
in $^{10}$Be and $^{10}$C. By analyzing the energies and the
overlaps between wave function components, we observe two different mechanisms
enhancing the formation of deuteron-like pairs in $^{10}$B. We also discuss the
pairing effect by showing average distances between components in each nucleus
and density distributions of valence nucleons.
\end{abstract}
\maketitle

\section{Introduction}
Investigating of nucleon-nucleon ($NN$) pairing effect is one of the most
interesting topics in nuclear structure theories \cite{En'yo2015, Morita2016}.
Especially, the knowledge of paring effects is essential for the understanding
of $NN$ correlations in light nuclei. Moreover, coupling of $\alpha$-clusters
and $NN$ pairs is important for the cluster states of general nuclei composed of
both $\alpha$-clusters and valence nucleons, as discussed in various many-body
systems in previous works \cite{Yamada2005, Funaki2015}. Hence, investigation of
the $NN$ pairing effect in cluster states is a meaningful step to improve our
present understanding of nuclear clustering effects \cite{Xu2006, Ren2012}.

There are two kinds of $NN$ pairs respecting to the isospin symmetry, including
the isovector pairs with $T=1$, and the isoscalar pairs with $T=0$. For general
nuclei, the isovector pairs are studied intensively but the deuteron-like
isoscalar pairs are relatively rare \cite{Frauendorf2014, Ropke2000}. One
important anomaly is the strong isoscalar pairing in $^{10}$B nucleus where the
formation of $NN$ pairs are strongly influenced by the coexistent
$\alpha$-clustering effect in this nucleus \cite{En'yo2015, Morita2016}.
Therefore, the mechanism for the formation of $NN$ pairing in $^{10}$B is
essential for the understanding of isoscalar pairing effect. Considering the
complexity that originate from the coupling of clustering and pairing effects,
it is desirable to fix the $\alpha$-cluster components in nuclei and then pin
down the modulation of $NN$ pairing by the spin-isospin channels of two paired
nucleons. An ideal approach is to compare the $NN$ pairing effects between
$^{10}$B nucleus in $T=0$ states with $^{10}$Be and $^{10}$C nuclei in $T=1$
states. Correspondingly, theoretical descriptions should be formulated for these
nuclei to investigate their $NN$ pairing structures and dynamics of pair
motion. Pairing strength obtained from these investigations is also essential
for experimental probing of $NN$ correlations where the theoretical predictions
of $(p, pd)$ nuclear reactions observables are included as inputs
\cite{Chazono2017}.

In order to study the different pairs consisting of neutrons ($nn$),
protons ($pp$), or proton and neutron ($pn$), we focus on the $^{10}$Be, $^{10}$B
and $^{10}$C nuclei that are all composed of two valence nucleons and two
$\alpha$-clusters. In our study, because of the different configurations of
nucleon-pairs among these three nuclei, it is possible to discuss the essential
mechanisms for the formation of $NN$ pairs as well as their different
properties, especially for the deuteron-like proton-neutron correlation in
$^{10}$B.

To study the pairing effects in these nuclei, we propose a new extended
formulation of the Tohsaki-Horiuchi-Schuck-R\"{o}pke (THSR) wave function, which
is a successful clustering model for various light nuclei \cite{Tohsaki2001,
Funaki2002, Zhou2012, Zhou2013}, especially for the Hoyle state ($0_2^+$) in
$^{12}$C \cite{Tohsaki2001}. By comparing with generator coordinate method (GCM),
the wave function of cluster states are found to be almost 100\% accurately
described by a single THSR wave function in light nuclei \cite{Funaki2003,
Funaki2005, Funaki2009, Zhou2013, Zhou2012}. In this work, we propose the
extended THSR wave function in superposed form, which is named as ``THSR+pair''
wave function. For the first time, additional pairing configuration for valence
nucleons is introduced to the THSR approach, which provides a convenient
framework for the discussion of the $NN$ pairing effect in nuclear system. With
this wave function, we investigate the abnormal isoscalar $NN$ paring effects
for the $^{10}$B nucleus and compare with the isovector $NN$ pairing effects in
${}^{10}$Be and ${}^{10}$C nuclei. Moreover, benefiting from the intrinsic
concision in analytical formulation, the THSR wave function possesses great
advantage in discussing the structure and dynamics of $NN$ pairs in nuclei, as
for the $\alpha$-clusters \cite{Zhou2012, Zhou2014} and valence nucleons
\cite{Lyu2015,Lyu2016,Lyu2017} in previous works.

This work is organized as following. In Section \ref{sec:waveFunction}, we
formulate the THSR+pair wave function for ${}^{10}$Be, ${}^{10}$B and ${}^{10}$C
nuclei. In Section \ref{sec:results} we provide the numerical results, including
the energy, distances between components, density distributions and
corresponding discussions. The last Section \ref{sec:conclusion} contains the
conclusions.

\section{Formulation}
\label{sec:waveFunction}
We start by writing the traditional THSR wave function which is used in our
previous calculations \cite{Lyu2015},
\begin{equation}\label{eq:lyu}
   \begin{split}
\Phi = \prod_{i=1}^2\int &dR_i{\rm exp}(-\frac{R_{i,x}^2}{\beta_{\alpha,xy}^2}-\frac{R_{i,y}^2}{\beta_{\alpha,xy}^2}-\frac{R_{i,z}^2}{\beta_{\alpha,z}^2})\\
&\times\int dR_a{\rm exp}(-\frac{R_{a,x}^2}{\beta^2_{a,xy}}-\frac{R_{a,y}^2}{\beta^2_{a,xy}}-\frac{R_{a,z}^2}{\beta^2_{a,z}})\\
&\times\int dR_b{\rm exp}(-\frac{R_{b,x}^2}{\beta^2_{b,xy}}-\frac{R_{b,y}^2}{\beta^2_{b,xy}}-\frac{R_{b,z}^2}{\beta^2_{b,z}})\\
&\times
e^{im_{a}\phi_{\mathbf{R}_{a}}}e^{im_{b}\phi_{\mathbf{R}_{b}}}\Phi^B(\mathbf{R}_{1},\mathbf{R}_{2},\mathbf{R}_{a},\mathbf{R}_{b}),
   \end{split}
\end{equation}
The Gaussian parameter $\beta$s constrain the nonlocalized motions of two
$\alpha$ clusters and valence nucleons \cite{Lyu2015}. We choose difference
parameter $\beta$s for $z$ direction and $x-y$ direction separately because of
the deformation in nuclei. These parameters are determined by variational
calculation. $\Phi^B$ is the Brink wave function, where $\mathbf{R}_{1,2}$ and
$\mathbf{R}_{a,b}$ are corresponding generate coordinates for the $\alpha$
clusters and valence nucleons. The terms $e^{im_{a}\phi_{\mathbf{R}_{a}}}$ and
$e^{im_{b}\phi_{\mathbf{R}_{b}}}$ are the phase factors which are introduced to
obtain correct parities and orbital angular momenta for valence
nucleons \cite{Lyu2015}.

The single-particle wave function can be written as
\begin{equation}
  \begin{split}
\phi(\mathbf{r}) = \left(\frac{2\nu}{\pi}\right)^{3/4}e^{-\nu(\mathbf{r}-\mathbf{R})^2}\chi_{\sigma}\chi_{\tau},
  \end{split}
\end{equation}
where the Gaussian range parameter $\nu = \frac{1}{2b^2}$. $\chi_{\sigma}$ is
the spin part of nucleon which is up ($\uparrow$) or down ($\downarrow$) in the
$z$-direction. $\chi_{\tau}$ is the isospin part of proton ($p$) or neutron
($n$).

The above traditional THSR wave function in Eq.~(\ref{eq:lyu})
provides good description for the molecular-orbit configurations but does not
include directly the descriptions for the $NN$ pairing structure. In pioneer
works, many studies indicate that the two valence nucleons have trends to
formulate $NN$ pair in $^{10}$Be, $^{10}$B and $^{10}$C nuclei
\cite{Kobayashi2013, En'yo2015}. In order to describe this component and provide
clear description for the pair structure, we introduce an additional compact
$NN$ pairing term as
\begin{equation}\label{eq:pair}
   \begin{split}
\Phi_{p} = \prod_{i=1}^2\int &dR_i{\rm exp}(-\frac{R_{i,x}^2}{\beta_{\alpha,xy}^2}-\frac{R_{i,y}^2}{\beta_{\alpha,xy}^2}-\frac{R_{i,z}^2}{\beta_{\alpha,z}^2})\\
&\times\int dR_{\rm pair}{\rm exp}(-\frac{R_{\rm pair,\it x}^2}{\beta^2_{\rm pair,\it xy}}-\frac{R_{\rm pair,\it y}^2}{\beta^2_{\rm pair,\it xy}}-\frac{R_{\rm pair,\it z}^2}{\beta^2_{\rm pair,\it z}})\\
&\times
e^{im_{a}\phi_{\mathbf{R}_{\rm pair}}}e^{im_{b}\phi_{\mathbf{R}_{\rm pair}}}\Phi^B(\mathbf{R}_{1},\mathbf{R}_{2},\mathbf{R}_{\rm pair}).
   \end{split}
\end{equation}
In this term, we treat valence nucleons as a two-particle cluster, which share
the same generate coordinate $\mathbf{R}_{\rm pair}$. This corresponds to a pairing
configuration in $^{10}$Be, $^{10}$B and $^{10}$C. We formulate the THSR+pair
wave function as a superposition of the molecular-orbit configuration $\Phi$ in
Eq.(1) and this additional term $\Phi_{p}$, as
\begin{equation}
\Psi_{\rm pair} = a\Phi+b\Phi_{p}.
\end{equation}
Here $a$ and $b$ are the coefficient parameters which are determined by
variational calculations. The parameters $m_a$ and $m_b$ in the phase factors
$e^{im_{a}\phi_{\mathbf{R}_{a}}}$ and $e^{im_{b}\phi_{\mathbf{R}_{b}}}$ are
chosen according to the rotational symmetry of the nuclear state under
consideration \cite{Lyu2015, Lyu2016}. For the $0^+$ ground state of $^{10}$Be
and $^{10}$C, we choose $m_a=1$ and $m_b=-1$ in Eqs.~(\ref{eq:lyu}) and
(\ref{eq:pair}) to describe the antiparallel couplings of spins for the two
valence nucleons around two $\alpha$ clusters. We note that under this condition
the phase factors in Eq.~(\ref{eq:pair}) vanish and the pair wave function
$\Phi_p$ is reduced to the $s$-wave. As for the $3^+$ ground state of $^{10}$B,
the parameters are chosen as $m_a=m_b=1$, which describes the parallel couplings
of spins between valence nucleons. 

We apply the angular-momentum projection technique $\hat{P}_{MK}^{J}\left| \Psi
\right\rangle$ to restore the rotational symmetry \cite{Schuck1980},
\begin{equation}
  \begin{split}
\left| \Psi^{JM} \right\rangle&=\hat{P}_{MK}^{J}\left| \Psi \right\rangle\\
    &=\frac{2J+1}{8\pi^{2}}\int d \Omega D^{J*}_{MK}(\Omega)\hat R (\Omega)
    \left| \Psi \right\rangle,
  \end{split}
\end{equation}
where $J$ is the total angular momentum of the system. For the $3_1^+0$ ground
state and $1_1^+0$ excited state of $^{10}$B with isospin $T=0$, we take the
isospin projection by using the proton-neutron exchange operator
$\hat{P}_{p{\leftrightarrow}n}$ as in introduced Refs.~\cite{En'yo2015,
Morita2016}.

The Hamiltonian of the $A=10$ nuclear systems can be written as
\begin{equation}\label{hamiltonian}
  H=\sum_{i=1}^{10} T_i-T_{c.m.} +\sum_{i<j}^{10}V^N_{ij}
    +\sum_{i<j}^{10}V^C_{ij} +\sum_{i<j}^{10}V^{ls}_{ij}.
\end{equation}
For the central force in $NN$ interaction, the Volkov No. 2 interaction
\cite{Volkov1965} is selected as
\begin{equation}\label{vn}
  V^N_{ij}=\{V_1 e^{-\alpha_1 r^2_{ij}}-V_2 e^{-\alpha_2 r^2_{ij}}\}
  \{ W - M \hat P_\sigma \hat P_\tau \ + B \hat P_{\sigma} - H \hat P_{\tau}\},
\end{equation}
where $M=0.6$, $W=0.4$, $B=H=0.125$, $V_{1}=-60.650$ MeV,
$V_{2}=61.140$ MeV, $\alpha_{1}=0.309$ fm${}^{-2}$, and $\alpha_{2}=0.980$
fm${}^{-2}$. The G3RS (Gaussian soft core potential with three ranges) term
\cite{Yamaguchi1979}, which is a two-body type interaction, is taken as the
spin-orbit interaction, as
\begin{equation}\label{vc}
V^{ls}_{ij}=V^{ls}_0\{
           e^{-\alpha_1 r^2_{ij}}-e^{-\alpha_2 r^2_{ij}}
           \} \mathbf{L}\cdot\mathbf{S} \hat{P}_{31},
\end{equation}
where $\hat P_{31}$ projects the two-body system into triplet odd state and the
parameters are set to be $V_{0}^{ls}$=1600 MeV, $\alpha_{1}$=5.00 fm${}^{-2}$
and $\alpha_{2}$=2.778 fm${}^{-2}$. The Gaussian width parameter $b$ of single
particle wave functions is chosen as $b=1.46$ fm.

\section{Results and Discussion}
\label{sec:results}

We calculate the ground state energies of $^{10}$Be, $^{10}$B and $^{10}$C by
variational optimization of parameters in the THSR+pair wave function. The
corresponding energy results are shown in Table \ref{table:result1}, where
corresponding experimental data \cite{Tilley2004} and results calculated with
the traditional THSR wave function $\Phi$ in Eq.~(\ref{eq:lyu}) are also
included. The masses of proton and neutron in the single-particle wave function
are set to be to experimental values in Ref.~\cite{Audi2012}.

\begin{table*}[htbp]
  \begin{center}
    \caption{\label{table:result1}
    Energies of the ground states for $^{10}$Be, $^{10}$B and $^{10}$C.
    $E^{\text{THSR}}$ denotes energies obtained from the THSR wave function,
    $E^{\text{THSR+pair}}$ denotes results obtained from the THSR+pair wave
    function. $\Delta$ denotes the improvement of energies in the new THSR+pair
    wave function comparing to the values obtained from previous version of THSR
    wave function. All units of energies are in MeV.}
 \begin{tabular*}{12cm}{ @{\extracolsep{\fill}} l c c c}
    \hline
    \hline
 &$^{10}$Be($0^+1$) &$^{10}$B($3^+0$) &$^{10}$C($0^+1$)\\
    \hline
$E^{\text{exp}}$\cite{Tilley2004}         &-65.0 &-64.8 &-60.3 \\
$E^{\text{THSR}}$        &-58.3 &-59.8 &-54.4 \\
$E^{\text{THSR+pair}}$   &-59.2 &-61.8 &-55.3 \\
$\Delta$                 & 0.9  & 2.0  & 0.9  \\
    \hline
    \hline
  \end{tabular*}
  \end{center}
\end{table*}

From the comparison in Table \ref{table:result1}, it is clearly observed that
the ground state energies of $^{10}$Be, $^{10}$B and $^{10}$C are greatly
improved by additional superposition of the pair term $\Phi_p$ in Eq.~(2) in the
THSR+pair wave function $\Psi_{\rm pair}$. These results prove that the $NN$ pairing
structure is crucial for precise descriptions of these nuclei. For $^{10}$B, the
THSR+pair wave function improves the ground state energy for about 2.0 MeV
comparing to traditional THSR wave function, which is more significant than the
improvements for other two nuclei. This indicates that the isoscalar $NN$
pairing effect with $T=0$ in $^{10}$B is stronger than those isovector
counterparts with $T=1$ in $^{10}$Be and $^{10}$C nuclei.

In order to investigate the $NN$ paring strength in the THSR+pair wave function,
we calculate the overlap between molecular-orbit term $\Phi$ and the total
THSR+pair wave function $\Phi_{\rm pair}$, the overlap between pairing term $\Phi_p$
and the total THSR+pair wave function $\Phi_{\rm pair}$, as well as the overlap
between molecular-orbit term $\Phi$ and pairing term $\Phi_p$. Corresponding
results are shown in Table \ref{table:overlap}. From this table, we observe that
the overlaps between the molecular-orbit term and THSR+pair wave function are
larger than 90\% for all of $^{10}$Be, $^{10}$B and $^{10}$C nuclei. These large
overlaps indicate that the molecular-orbit term could provide good description
for these nuclei. However, additional pairing term is still necessary to obtain
accurate wave function as these overlaps do not equal 100 \%. It is also
observed that the overlaps $<\Phi_p|\Psi_{\rm pair}>^2$ are different among
$^{10}$Be, $^{10}$B and $^{10}$C, where the overlap for $^{10}$B nucleus is
significantly larger than other two nuclei. Hence the optimized THSR+pair wave
describes stronger $NN$ pairing effect in $^{10}$B than those in ${}^{10}$Be and
${}^{10}$C as we concluded previously. From the giant ratio
$<\Phi|\Psi_p>^2$=75.8 \%, we found that molecular-orbit term $\Phi$ in the
$3^+0$ ground state of $^{10}$B provides the description that is analogous to
the pairing term $\Phi_{p}$, which explains the strong pairing effect in this
state.

\begin{table*}[htbp]
  \begin{center}
    \caption{\label{table:overlap} The overlaps between each two of the
    THSR+pair wave function $\Psi_{\rm pair}$ and its two components $\Phi$ and
    $\Phi_{p}$ of molecular-orbit configuration and pairing configuration,
    respectively. Values are calculated for the ground states of $^{10}$Be,
    $^{10}$B and $^{10}$C nuclei. All the wave functions $\Phi$, $\Phi_p$ and
    $\Psi_{\rm pair}$ have been normalized.}
 \begin{tabular*}{10cm}{ @{\extracolsep{\fill}} c c c c }
    \hline
    \hline
 &$^{10}$Be($0^+1$) &$^{10}$B($3^+0$) &$^{10}$C($0^+1$)\\
    \hline
$<\Phi|\Psi_{\rm pair}>^2$          &92.9\% &90.8\% &93.6\%\\
$<\Phi_p|\Psi_{\rm pair}>^2$        &56.6\% &93.3\% &67.5\% \\
$<\Phi|\Psi_p>^2$          		&45.5\% &75.8\% &43.1\%\\
    \hline
    \hline
  \end{tabular*}
  \end{center}
\end{table*}

We also calculate the $1_1^+0$ excited state of $^{10}$B, which is the
counterpart for the $0_1^+$ ground states of $^{10}$Be and $^{10}$C nuclei, as
they all have dominant $L = 0$ components that originate from the antiparallel
coupling of orbital angular momentum of two valence nucleons \cite{En'yo2015,
Morita2016}. Hence, the total spin $S$ and isospin $T$ of two valance nucleons
are the only differences among the $1_1^+0$ excited state of $^{10}$B and the
$0_1^+$ ground states of $^{10}$Be and $^{10}$C. In Table
\ref{table:excitedstate}, we show the energies of the $1_1^+$ excited state
calculated with the traditional THSR wave function $\Phi$ and the THSR+pair wave
function $\Psi_{\rm pair}$. In these calculations, we set parameters $\text{m}_{ab}
= \pm1$ respectively for two valence nucleons to describe the antiparallel
coupling of orbital angular momenta. The corresponding experimental data adopted
from Ref.~\cite{Tilley2004} are also included for comparison. From this table,
it is observed that the excitation energy of the $1_1^+0$ state is improved from
2.8 MeV to 1.0 MeV by adding the pairing term, which is much closer to the
experimental value 0.7 MeV. It is also clearly shown that the introduction of
additional pairing term $\Phi_p$ improves energy of the $1_1^+0$ excited state
by about 3.8 MeV, which is significantly larger than the corresponding
improvement of about 2.0 MeV for the $3^+0$ ground state. This drastic
difference indicates even stronger $NN$ pairing effect in the $1_1^+0$ excited
state. 

Furthermore, this improvement of 3.8 MeV is significantly larger
than corresponding values for the $0_1^+1$ ground states of $^{10}$Be and
$^{10}$C, as shown in Table \ref{table:result1}. Hence we observe the enhanced
pairing effect again in the $1^+_10$ state of $^{10}$B with $T=0$ comparing to
the pairs with $T=1$ in $^{10}$Be and $^{10}$C. However, the previous
explanation for the pairing effect in the $3^+0$ ground state of $^{10}$B no
longer persists, because the analogy between the molecular-orbit configuration
and pairing configuration is much weaker in the $1_1^+0$ state. This is
demonstrated in Table \ref{table:overlap1} where the overlap between the
molecular-orbit term $\Phi$ and the pairing term $\Phi_p$ in this state is found
to be 55.8 \%, which is much smaller than the corresponding value of 75.8 \% in
the $3^+0$ ground state. From this we conclude that these two configurations
compete with each other in the $1_1^+0$ state, which is different from the
analogous contribution from these two configurations in the total wave function.

As listed in Table \ref{table:overlap1}, the squared overlaps between the
molecular-orbit term $\Psi$ and total wave function $\Psi_{\rm pair}$ in the $T=1$
states of $^{10}$Be and $^{10}$C are larger than 90\%, which shows that the
molecular-orbit configuration prevails in these states. This can be explaind by
the fact that the molecular-orbit configuration is energetically favorable in
$T=1$ nuclei $^{10}$Be and $^{10}$C, where both the molecular orbits have
parallel spin-orbit coupling and provide large contributions to the total
energies of nuclei. In contradiction, the spin-orbit contributions from valance
nucleons is cancelled by each other in the pairing configuration, as the two
paired nucleons have opposite spin directions but the same orbital motion. In
the $1_1^+0$ excited state of $^{10}$B the dominance of molecular-orbit
configuration disappears, as shown by the larger overlap 87.5 \% between pairing
configuration $\Phi_{p}$ and total wave function $\Psi_{\rm pair}$. In the
molecular-orbit term $\Phi$ of this state, the spin-orbit coupling is parallel
and antiparallel for the two valance nucleons, respectively. Hence there is also
cancellation of spin-orbit contribution from valance nucleons, which is similar
to the case in pairing configuration. As a consequence, the molecular-orbit
configuration is not energetically favorable and it is quenched in the total
wave function by its competition with pairing configuration in the $1_1^+0$
excited state of $^{10}$B.

Here we have observed that the $NN$ pairing is formulated in
deuteron like channels ($T=0, S=1$) of two valance nucleons both the $3^+0$
ground state and the $1^+0$ excited state of $^{10}$B. However, the $NN$ pairs
are formulated in two different mechanisms in these states. For the $3^+0$
ground state, the molecular-orbit configuration of two valance nucleons is
analogous to the pairing configuration and hence enhances the possibility of
$NN$ pairing. In the $1^+0$ excited state of $^{10}$B, molecular-orbit
configuration competes with the pairing configuration and hence is quenched
comparing to its dominance in $0_1^+1$ states of $^{10}$Be and $^{10}$C, which on
the other hand encourages the formation of $NN$ pairing.

\begin{table*}[htbp]
  \begin{center}
    \caption{\label{table:excitedstate}
    Energies of the $3^+0$ ground state and the $1_1^+0$ excited state for
    $^{10}$B. $E^{\text{THSR}}$ denotes energies obtained from the THSR wave
    function. $E^{\text{THSR+pair}}$ denotes results obtained from the THSR+pair
    wave function. $\Delta$ denotes the improvement of energies in the new
    THSR+pair wave function comparing to the values obtained from traditional
    THSR wave function. $E^{\text{exp}}$ denotes experimental values adopted from
    Ref.~\cite{Tilley2004}. $E_{\text{ex}}$ denotes the corresponding excited
    energies. All units of energies are in MeV.}
 \begin{tabular*}{8cm}{ @{\extracolsep{\fill}} c c p{1.5cm}|p{1cm}}
    \hline
    \hline
$^{10}$B &$3^+0$ &$1_1^+0$ &$E_{\text{ex}}$\\
    \hline
$E^{\text{exp}}$         &-64.8 &-64.1 &0.7 \\
$E^{\text{THSR}}$        &-59.8 &-57.0 &2.8 \\
$E^{\text{THSR+pair}}$   &-61.8 &-60.8 &1.0 \\
$\Delta$   				&2.0 &3.8 &1.8 \\
    \hline
    \hline
  \end{tabular*}
  \end{center}
\end{table*}

\begin{table*}[htbp]
  \begin{center}
    \caption{\label{table:overlap1}
    The overlaps between each two of the THSR+pair wave function $\Psi_{\rm pair}$
    and its two components $\Phi$ and $\Phi_{p}$ of molecular-orbit
    configuration and pairing configuration, respectively. Values are calculated
    for the ground states of $^{10}$Be and $^{10}$C nuclei and the $1_1^+0$
    excited state of $^{10}$B. All the wave functions $\Phi$, $\Phi_p$ and
    $\Psi_{\rm pair}$ have been normalized.}
 \begin{tabular*}{10cm}{ @{\extracolsep{\fill}} c c c c}
    \hline
    \hline
 &$^{10}$Be($0^+1$) &$^{10}$C($0^+1$) &$^{10}$B($1^+0$)\\
    \hline
$<\Phi|\Psi_{\rm pair}>^2$          &92.9\% &93.6\% &83.7\%\\
$<\Phi_p|\Psi_{\rm pair}>^2$        &56.6\% &67.5\% &87.5\%\\
$<\Phi|\Psi_p>^2$          		&45.5\% &43.1\% &55.8\%\\
    \hline
    \hline
  \end{tabular*}
  \end{center}
\end{table*}

\begin{figure}[htbp]
  \centering
  \includegraphics[width=0.60\textwidth]{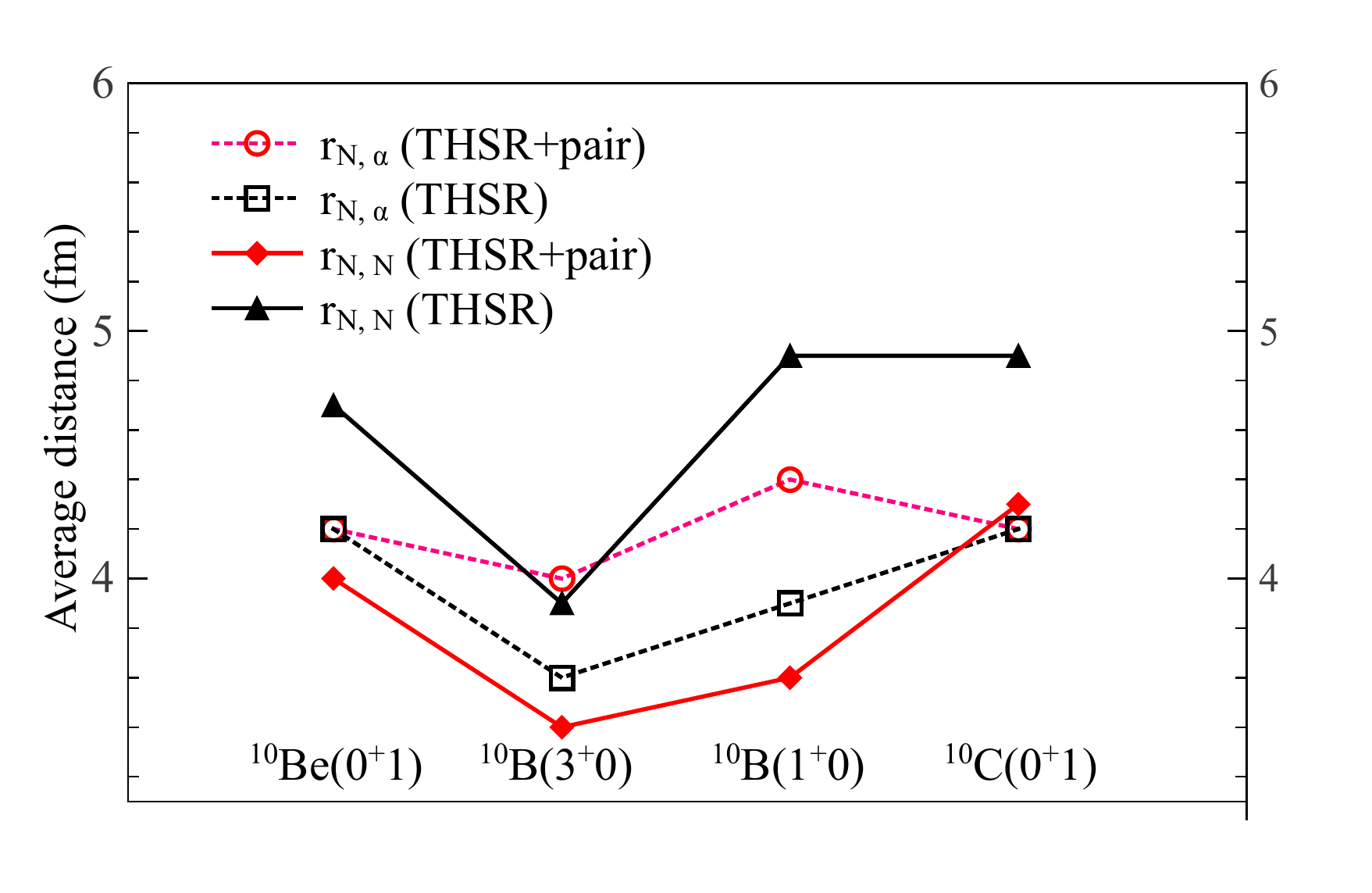}
  \caption{\label{fig:distance}
  The average distances between nucleons in $^{10}$Be($0^+1$), $^{10}$B($3^+0$),
  $^{10}$B($1^+0$) and $^{10}$C($0^+1$) states. The solid lines denote the
  average distances $r_{N,N}$ between two valence nucleons. The dashed lines
  denote the average distances $r_{N,\alpha}$ between valence nucleons and the
  center of two $\alpha$-clusters. For both the solid lines and dashed lines,
  the black color denotes results obtained by using only the molecular-orbit
  configuration $\Phi$ and the red color denotes results from the total
  THSR+pair wave function $\Phi_{\rm pair}$. All units are in fm.}
\end{figure}

The $NN$ pair structure in $^{10}$Be, $^{10}$B and $^{10}$C can be demonstrated
explicitly by showing the average distances between the two valence nucleons,
which correspond to the average sizes of $NN$ pairs. It should be noticed that
the formation of $NN$ pairs can affect both the distance between two valence
nucleons $r_{N,N}$ and the distance between $NN$ pair and the center of two
$\alpha$-clusters $r_{N,\alpha}$. Hence when the strength of $NN$ pairing effect
increases, the ratio $r_{N,N}/r_{N,\alpha}$ is reduced to a relatively small
value. We compare the average distances for $^{10}$Be, $^{10}$B and $^{10}$C in
Fig. \ref{fig:distance}, including the $1^+_10$ state of $^{10}$B. As shown in
this figure, with the molecular-orbit configurations (the black lines), the $NN$
distances $r_{N,N}$ have almost the same magnitude as $r_{N,\alpha}$ for all
nuclei of $^{10}$Be, $^{10}$B and $^{10}$C, which is due to the missing of
paring term in the molecular-orbit configuration. With the new extended
THSR+pair wave function (the red lines), the $r_{N,N}$ in $^{10}$Be is smaller
but comparable to $r_{N,\alpha}$, showing a relatively weak $NN$ pairing effect
in $^{10}$Be. For ${}^{10}$C, the $NN$ pairing effect is even weaker as
corresponding $r_{N,N}$ is larger than $r_{N,\alpha}$. We see significantly
small ratio $r_{N,N}/r_{N,\alpha}$ for both states of ${}^{10}$B nucleus, where
the $NN$ pairing effect is stronger, as discussed previously.

\begin{figure}[htbp]
  \centering
  \includegraphics[width=0.35\textwidth]{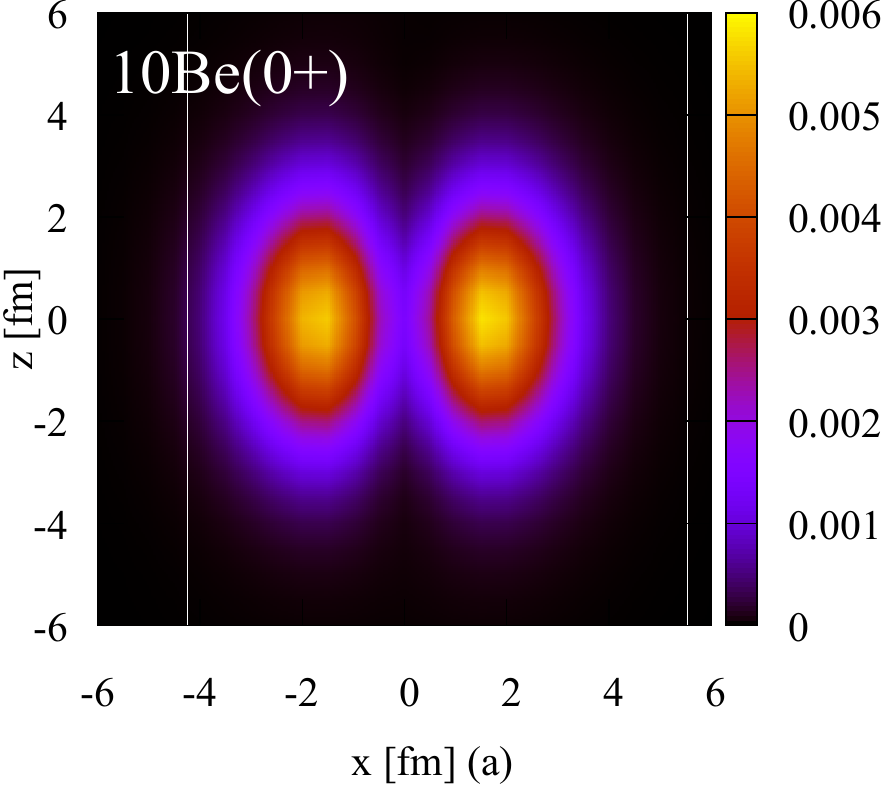}
  \includegraphics[width=0.35\textwidth]{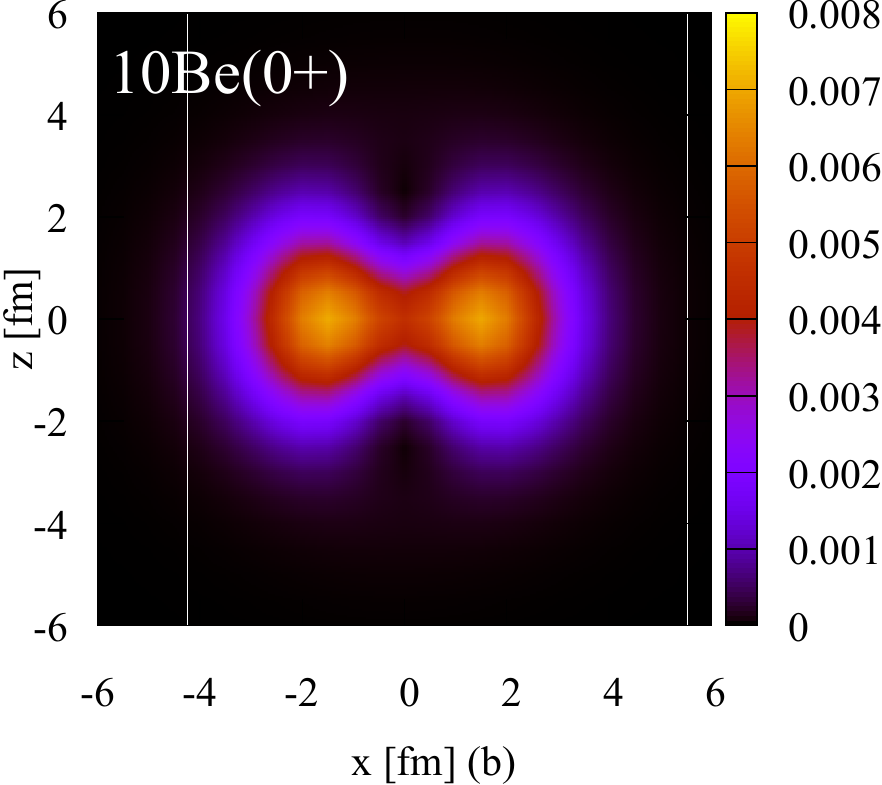}
  
  \includegraphics[width=0.35\textwidth]{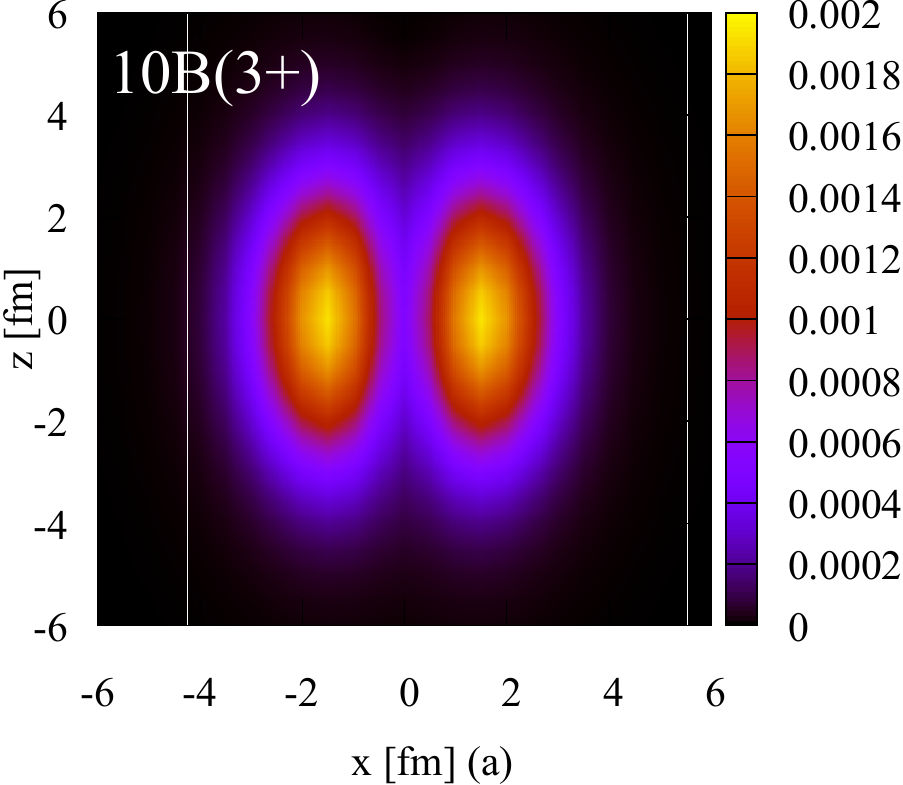}
  \includegraphics[width=0.35\textwidth]{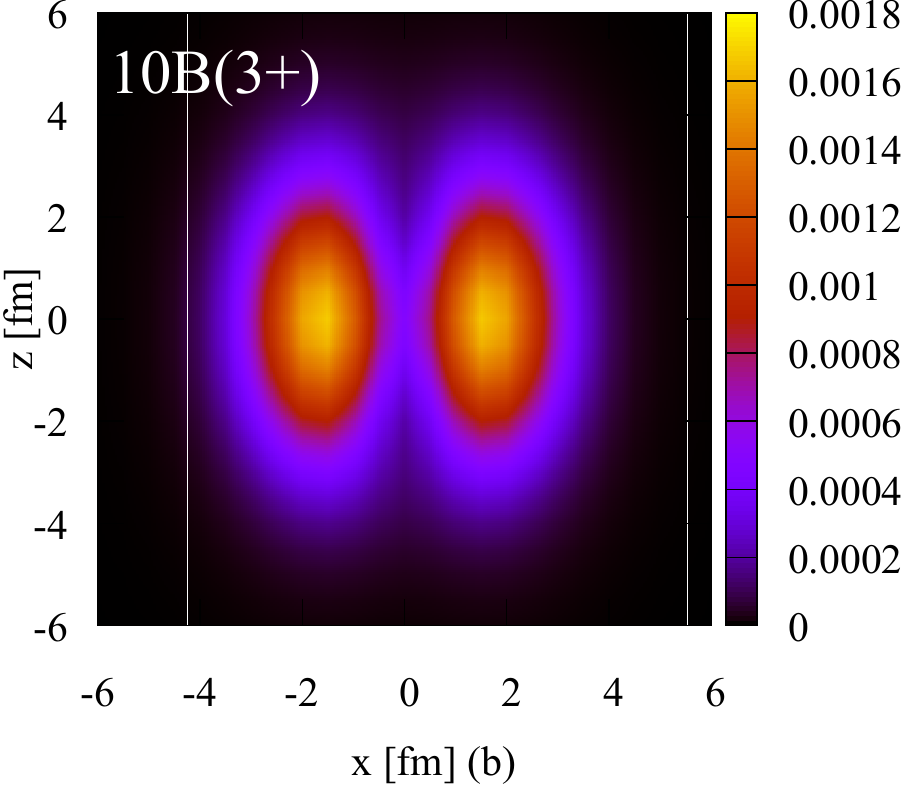}
  
  \includegraphics[width=0.35\textwidth]{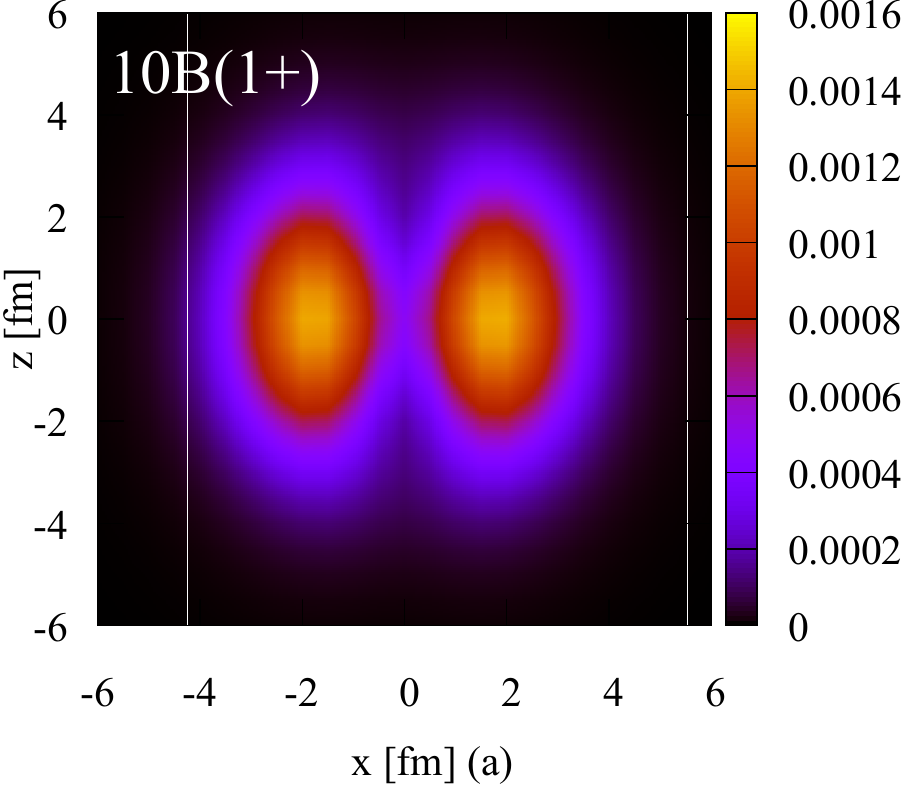}
  \includegraphics[width=0.35\textwidth]{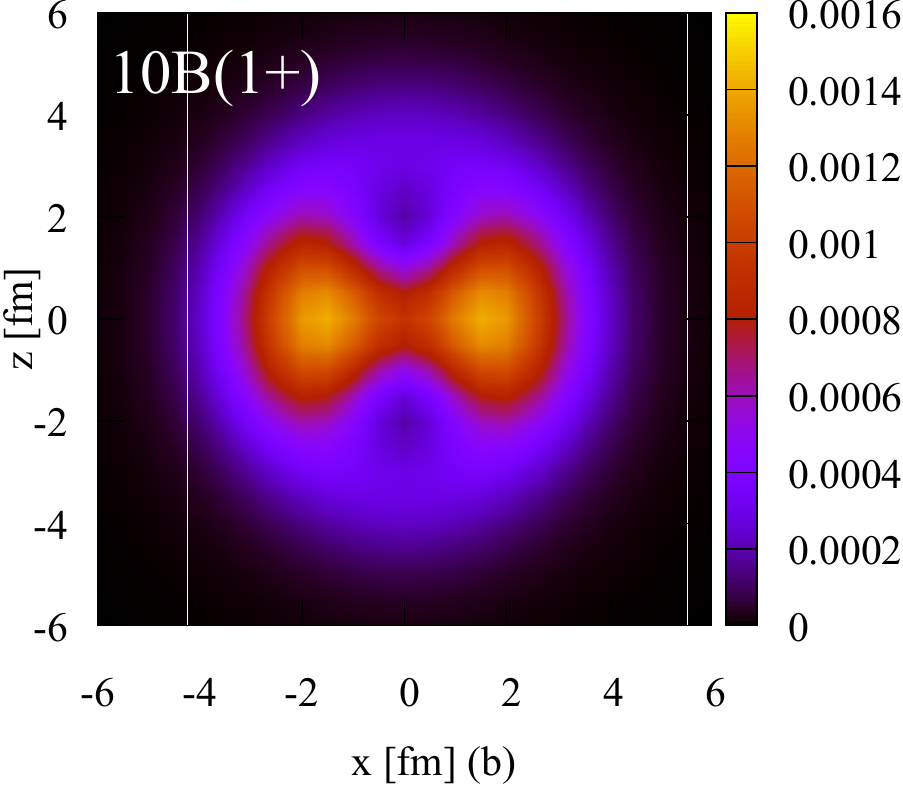}
  \caption{\label{fig:densityExtra}
	The density distributions of the valance nucleons on $x-z$ plane for the
$0^+1$ state of ${}^{10}$Be, the $3^+0$ state of ${}^{10}$B and the $1^+0$ state
of ${}^{10}$B. The panels (a) are calculated with the THSR+pair wave function
$\Phi_{\rm pair}$ with optimized parameters. The panels (b) are obtained by using
only the pairing term $\Phi_p$ with parameters $a=0$. For all these
calculations, $\beta$ parameters are set to be optimized values in corresponding
THSR+pair wave functions.}
\end{figure}

In order to investigate the nuclear dynamics of $NN$ pairs, we calculate the
density distributions for valence nucleons of the $0^+1$ ground state of
$^{10}$Be, the $3^+0$ ground state of $^{10}$B and $1^+_10$ excited state of
$^{10}$B, as shown in Fig. \ref{fig:densityExtra}. In this figure, panels
labeled by ``(a)" are calculated with the THSR+pair wave function $\Phi_{\rm pair}$.
The panels labeled by ``(b)'' are calculated by using only the pairing term
$\Phi_p$. For all these wave functions, $\beta$ parameters are set to be
optimized values in corresponding THSR+pair wave functions.

From this figure, we observe that the valence nucleons in the $3^+0$ ground
state of $^{10}$B have narrow distributions in the $x-y$ direction because it is
tightly bounded by the spin-orbit potential and the centrifugal barrier. While
in the cases of the ground state of $^{10}$Be and the $1^+0$ excited state of
$^{10}$B, the distributions of the valence nucleons are more broad because of
the weaker spin-orbit potential and lower centrifugal barrier. This result
agrees with the conclusions in Ref.~\cite{Morita2016} and Ref.~\cite{Morita2017}. By comparing the ``(b)'' panels in these figures, we notice
that when no centrifugal barrier exists,  as in the $0^+1$ state of $^{10}$Be
and $1^+0$ state of $^{10}$B, the $NN$ pairs described by the pairing term
$\Psi_p$ are likely to have more distribution near $z=0$ cross section between
two $\alpha$ clusters , which corresponds to a relatively dilute three-clusters
structure of $\alpha$+$\alpha$+pair. On the other hand, in the $3^+0$ ground
state of $^{10}$B, the strong spin-orbit coupling from the orbital angular
momentum $L=2$  encourages the spreading of valance nucleons in the
$z$-direction around $\alpha$-clusters to formulate $\pi$-molecular orbits, as
discussed in Refs.~\cite{Lyu2015,Lyu2016}. 

The similar conclusions also can be made depending on the optimum parameters of the THSR+pair wave function or
traditional THSR wave function, which are listed in Table \ref{table:parameter} for
each state.

\begin{table*}[htbp]
  \begin{center}
    \caption{\label{table:parameter}The variationally optimized $\beta$
    parameters for the wave function of $^{10}$Be ($0^+$), $^{10}$B ($3^+$),
    $^{10}$B ($1^+$) and $^{10}$C ($0^+$). For each nucleus, the upper line
    corresponds to the calculation with traditional THSR wave function $\Phi$
    and the lower line denoted by symbol ``$\hookrightarrow$" corresponds to
    calculation with THSR+pair wave function $\Phi_{\rm pair}$. The units of $\beta$
    parameters are in fm.}
 \begin{tabular*}{13cm}{@{\extracolsep{\fill}} 
    c|p{1cm}p{1cm}|p{1cm}p{1cm}|p{1cm}p{1cm}|p{1cm}p{1cm}}
    \hline
    \hline
Nucleus& $\beta_{\alpha,xy}$ &$\beta_{\alpha,z}$ &$\beta_{ab,xy}$
&$\beta_{ab,z}$ &$\beta_{\rm pair,xy}$ &$\beta_{\rm pair,z}$ &$a$ &$b$\\
    \hline
$^{10}$Be ($0^+$)  &0.1 &2.5 &1.9 &2.9 &/   &/   &/    &/ \\
$\hookrightarrow$  &0.1 &2.8 &1.9 &3.3 &2.5 &0.8 &0.89 &0.11 \\
\hline
$^{10}$B ($3^+$)   &0.1 &1.9 &1.1 &2.2 &/   &/   &/    &/ \\
$\hookrightarrow$  &0.1 &2.6 &1.0 &3.2 &1.8 &3.3 &0.77 &0.23 \\
\hline
$^{10}$B ($1^+$)   &0.1 &2.2 &2.2 &2.0 &/   &/   &/    &/ \\
$\hookrightarrow$  &0.1 &3.0 &2.4 &2.8 &3.0 &2.7 &0.77 &0.23 \\
\hline
$^{10}$C ($0^+$)   &0.1 &2.8 &2.2 &3.3 &/   &/   &/    &/ \\
$\hookrightarrow$  &0.1 &3.0 &2.4 &3.2 &2.5 &0.2 &0.89 &0.11 \\
    \hline
    \hline
  \end{tabular*}
  \end{center}
\end{table*}

\section{Conclusion}
\label{sec:conclusion}
We propose a new extended formulation of THSR wave function, named as
``THSR+pair" wave function, for $^{10}$Be, $^{10}$B and $^{10}$C nuclei. In this
wave function, $NN$ pairing term is introduced in addition to the
molecular-orbit term used in previous version of THSR wave function. By using
the THSR+pair wave function, the energies for the ground states are improved
significantly for these nuclei, especially the $^{10}$B nucleus. Analyses of
energies and overlaps show that the pair configuration is stronger in the
$^{10}$B nucleus comparing to other two nuclei. We also calculate the $1_1^+0$
excited state of $^{10}$B using the THSR+pair wave functions and observe again
strong pairing effect in this state. These results show that paring effect are
enhanced by the deuteron-like channel of $S = 1$ and $T = 0$ (isoscalar).
From the energies and overlaps between wave function components, we
found that there are two different mechanisms that enhance the formation of $NN$
pair in $^{10}$B nucleus. In the $3_1^+0$ ground state, the strong pairing
effect originates from the analogy between molecular-orbit configuration and
pairing configuration. In the $1_1^+0$ excited state, the pairing configuration
competes with molecular-orbit configuration and the molecular-orbit term is
energetically unfavored and quenched. We also discuss the structure of $NN$
pairs and their dynamics of motion in space, by calculating the average
distances between components in nucleus and the density distributions of valance
nucleons. This study further improves the understanding of the formation of $NN$
pairs and their properties, especially for those in isoscalar channels, which
could be beneficial for future investigations of $NN$ correlations and general
cluster states composed of both $\alpha$-clusters and $NN$ pairs.

\begin{acknowledgments}
The author would like to thank Dr. Wan for fruitful discussions. This work is
supported by the National Natural Science Foundation of China (grant nos
11535004, 11375086, 11120101005, 11175085, 11235001, 11761161001), by the
National Major State Basic Research and Development of China, grant no.
2016YFE0129300, by the Science and Technology Development Fund of Macao under
grant no. 068/2011/A, and by the JSPS KAKENHI GRANTS No.JP16K05351.
\end{acknowledgments}



\begin{thebibliography}{}
\bibitem{En'yo2015}Y. Kanada-En’yo and H. Morita, F. Kobayashi, Physical Review C \textbf{91}, 054323 (2015).
\bibitem{Morita2016}H. Morita and Y. Kanada-En'yo, Prog. Theor. Exp. Phys. 103D02 (2016).
\bibitem{Yamada2005}T. Yamada and P. Schuck, Eur. Phys. J. A
  \textbf{26}, 185 (2005).
\bibitem{Funaki2015}Y. Funaki, H. Horiuchi, A. Tohsaki, Progress in Particle and Nuclear Physics \textbf{82} (2015) 78–132.
\bibitem{Xu2006}Chang Xu and Zhongzhou Ren, Phys. Rev. C \textbf{73}, 041301 (2006).
\bibitem{Ren2012}Yuejiao Ren and Xhongzhou Ren, Phys. Rev. Lett. \textbf{85}, 044608 (2012)
\bibitem{Frauendorf2014}S. Frauendorf and A. O. Macchiavelli, Prog. Part. Nucl. Phys. \textbf{78}, 24 (2014).
\bibitem{Ropke2000}G. Ropke, A. Schnell, P. Schuck, and U. Lombardo, Phys. Rev. C \textbf{61}, 024306 (2000).
\bibitem{Kobayashi2013}F. Kobayashi and Y. Kanada-En'yo, arXiv:1312.0052 [nucl-th] (2013).
\bibitem{Chazono2017}Chazono, \textit{private communication} (2017).
\bibitem{Tohsaki2001}A. Tohsaki, H. Horiuchi, P. Schuck, and G. R\"{o}pke, Phys. Rev. Lett. \textbf{87}, 192501 (2001).
\bibitem{Funaki2002}Y. Funaki, H. Horiuchi, A. Tohsaki, P. Schuck, and G. R\"{o}pke, Prog. Theor. Phys. \textbf{108}, 297 (2002).
\bibitem{Zhou2013}B. Zhou, Y. Funaki, H. Horiuchi, Z. Ren, G. R\"{o}pke, P. Schuck, A. Tohsaki, C. Xu, and T. Yamada, Phys. Rev. Lett. \textbf{110}, 262501 (2013).
\bibitem{Zhou2012}B. Zhou, Z. Ren, C. Xu, Y. Funaki, T. Yamada, A. Tohsaki, H. Horiuchi, P. Schuck, and G. R\"{o}pke, Phys. Rev. C \textbf{86}, 014301 (2012).
\bibitem{Funaki2003}Y. Funaki, A. Tohsaki, H. Horiuchi, P. Schuck, G. 	R\"{o}pke, Phys. Rev. C \textbf{67}, 051306(R) (2003).
\bibitem{Funaki2005}Y. Funaki, A. Tohsaki, H. Horiuchi, P. Schuck, G. 	R\"{o}pke, Eur. Phys. J. A \textbf{24}, 321-342 (2005).
\bibitem{Funaki2009}Y. Funaki, H. Horiuchi, W. von Oertzen, G. R\"{o}pke, P. Schuck, A. Tohsaki and T. Yamada, Phys. Rev. C \textbf{80}, 064326 (2009).
\bibitem{Zhou2014}B. Zhou, Y. Funaki, H. Horiuchi, Z. Ren, G. R\"{o}pke, P. Schuck, A. Tohsaki, C. Xu, and T. Yamada, Phys. Rev. C \textbf{89}, 034319 (2014).
\bibitem{Lyu2015}M. Lyu, Z. Ren, B. Zhou, Y. Funaki, H. Horiuchi, G. R\"{o}pke, P. Schuck, A. Tohsaki, C. Xu, and T. Yamada, Phys. Rev. C \textbf{91}, 014313 (2015).
\bibitem{Lyu2016}M. Lyu, Z. Ren, B. Zhou, Y. Funaki, H. Horiuchi, G. R\"{o}pke, P. Schuck, A. Tohsaki, C. Xu, and T. Yamada, Phys. Rev. C \textbf{93}, 054308 (2016).
\bibitem{Lyu2017}M. Lyu, Z. Ren, H. Horiuchi, B. Zhou, Y. Funaki, G. R\"{o}pke, P. Schuck, A. Tohsaki, C. Xu, and T. Yamada, arXiv:1706.06538 [Nucl-Th] (2017).
\bibitem{Schuck1980} P. Ring and P. Schuck, \textit{The Nuclear Many-Body Problem} (Springer-Verlag, New York, 1980), p. 474.
\bibitem{Volkov1965}A.B. Volkov, Nucl. Phys. \textbf{74}, 33 (1965).
\bibitem{Yamaguchi1979}N. Yamaguchi, T. Kasahara, S. Nagata, and Y. Akaishi, Prog. Theor. Phys. \textbf{62}, 1018 (1979).
\bibitem{Tilley2004}D. R. Tilley, J. H. Kelley, J. L. Godwin, D. J. Millener, J. E. Purcell, C. G. Sheu, and H. R. Weller, Nucl. Phys. A \textbf{745}, 155 (2004).
\bibitem{Audi2012}G. Audi, F. G. Kondev, M. Wang, B. Pfeiffer, X. Sun, J. Blachot, and M. MacCormick, Chin. Phys. C \textbf{36}, 1157 (2012).
\bibitem{Navratil2007}P. Navratil, V. G. Gueorguiev, J. P. Vary, W. E. Ormand, and A. Nogga, Phys. Rev. Lett. \textbf{99}, 042501 (2007).
\bibitem{Navratil2002}P. Navratil and W. E. Ormand, Phys. Rev. C \textbf{68}, 034305 (2003).
\bibitem{Morita2017}H. Morita and Y. Kanada-En'yo, Phys.Rev. C \textbf{96}, 044318 (2017).
\bibitem{Funaki2012}Y. Funaki, T. Yamada, H. Horiuchi, G. R\"{o}pke, P. Schuck, and A. Tohsaki, Phys.Rev. Lett. \textbf{101}, 082502 (2012).
\end{thebibliography}
\end{document}